\documentclass[prb,showpacs,twocolumn]{revtex4}
\usepackage{amsmath}
\usepackage{graphicx}
\newcount\minute
\newcount\hour
\newcount\hourMins
\def\now
{
  \minute=\time
  \hour=\time \divide \hour by 60
  \hourMins=\hour \multiply\hourMins by 60
  \advance\minute by -\hourMins
  \zeroPadTwo{\the\hour}:\zeroPadTwo{\the\minute}
}
\def\timestamp
{
  \today\ \now
}
\def\today
{
  \zeroPadTwo{\the\day}-\zeroPadTwo{\the\month}-\the\year%
}
\def\zeroPadTwo#1%
{
  \ifnum #1<10 0\fi
  #1%
}
\def \dif {\mathrm{d}}

\pacs{74.81.-g,75.70.-i,74.25.Dw,74.78.-w}
\date{\timestamp}

\begin{document}

\title{Ground state configurations of vortices in superconducting film with magnetic dot}

\author{Zoran Ristivojevic}

\affiliation{Institut f\"ur Theoretische Physik,
Universit\"at zu K\"oln, Z\"ulpicher Stra\ss e 77,
D-50937 K\"oln, Germany}

\begin{abstract}
We consider a thin superconducting film with a magnetic
dot with permanent magnetization (normal to the film) placed on it by a method
based on London-Maxwell equations. For sufficiently high
dot magnetization a single vortex appears in the ground
state. Further increase of magnetization is accompanied
with the appearance of antivortices and more vortices in the film. We study analytically conditions for the
appearance of a vortex--antivortex pair for a range of
parameters. The phase diagram with diversity of
vortex--antivortex states is calculated numerically.
When appear in the ground state, antivortices are at
distances comparable to the dot radius. For not too large
dot radii the total vorticity in the ground state is
predominantly zero or one. Magnetic field due to the dot and vortices everywhere in space is calculated analytically.
\end{abstract}
\maketitle

\section{Introduction}

Type--II superconductors in an external homogeneous magnetic field are much studied and well understood.\cite{Bla+94,Tinkham} They accommodate
regularly distributed flux tubes--vortices for fields between the lower and the upper critical field.\cite{Abr57} Similar situation
occurs in type--II superconducting films in
perpendicular fields.\cite{Pea64,Fet+67} A difference is that the first critical field for films vanishes for macroscopically large samples.\cite{Fet+67,Fis80}

An interesting class of systems which have attracted a great attention only recently are ferromagnet--superconductor hybrid systems. There one examines the influence of a material with heterogeneous magnetization on a superconductor.\cite{Lyu+05} Direct contact between the magnetic material and the superconductor is avoided usually by a thin insulating
layer which suppresses the proximity effect.
These hybrid systems can be fabricated fully controlling their parameters.\cite{Vel+08} Inhomogeneous magnetization of ferromagnets generates magnetic field that penetrates the superconductor. As a response to that field, supercurrents (and vortices under certain conditions) are induced in the film. The magnetic field of supercurrents interacts with the magnetic subsystem. Therefore by tuning parameters of the magnetic subsystem we examine different phenomena of the composite system. Hybrid systems offer a number of realizations of new interesting phenomena which include pinning of magnetization induced
vortices, commensurability effects (between external
magnetic field and periodic structure of magnetic
material) on the superconductor resistivity and
others.\cite{Mar+97,Mor+98,Lyu+05,Vel+08}

In this paper we are focused on the simplest ferromagnet--superconductor hybrid system which consists of a single magnetic dot grown on top of a
type--II superconducting film. The magnetic dot is assumed to have
permanent magnetization normal to the film surface.
Despite its simplicity this system deserves attention
since a diversity of different ground states can be
realized in the parameter space. Dots with sufficiently small magnetization
produce only supercurrents in the film, while with
the increase of magnetization different
configurations of vortex states appear in the
ground state(GS).\cite{Mar+96,Lyu+98,Sas+00,Erd+02,Erd05b,Mil+03} This is in striking contrast with respect to films in homogeneous fields.\cite{Fet+67,Fis80} However, we expect that for large dot radii any magnetization induces vortices in the film, like the case with homogeneous magnetic field.

The main goal in this paper is to define regions in the parameter space with different ground state configurations of vortices. A possible question that is under debate in literature is whether antivortices may also be induced in the film. One finds different statements about the presence of antivortices: while some authors claim its existence,\cite{Sas+00,Mil+02,Mil+03} the others
do not find it.\cite{Erd+02,Erd05b} We think that rough
estimates in Ref.~\cite{Sas+00}, calculations
with magnetic dipole in Ref.~\cite{Mil+02} or
study inside the nonlinear Ginzburg--Landau
theory with restriction to zero total vorticity
states done in Ref.~\cite{Mil+03} are
insufficient, and offer an independent and
detailed study of this problem. Under which
conditions and where vortices and antivortices
appear will be answered in our paper.

Magnetization of the dot (which is normal to the film surface) produces magnetic field in the film, which is under the dot parallel to the magnetization, but antiparallel for larger distances than the dot radius. The magnetic field under the dot favors the appearance of vortices under the dot, the appearance of antivortices outside the dot. Whether some vortex--antivortex configuration have the smallest energy or not depends on details of the interaction energy.

Vortices have spatial structure and
the superconducting order parameter vanishes
roughly at distances smaller than $\xi$ around
the center of the vortex. Here $\xi$ denotes the
coherence length of a superconductor.\cite{Tinkham}
The nonlinear Ginzburg--Landau approach takes
this into account, but apart from numerical
treatment one can hardly get analytic
results.\cite{Mil+03} Our approach is based on
London--Maxwell equations.\cite{Erd+02,Erd05a}
Despite its simplicity (it treats vortices as
point objects) it an useful approach since one
may get analytic expressions for relevant
quantities which are asymptotically exact when
all lengths in the problem are larger than the
spatial vortex extension.

It is interesting to mention that the vortex nucleation in
superconducting microtriangles and squares occurs in such
a way that the symmetry of the superconductor is
preserved.\cite{Chibotaru+00,Chibotaru+01} For example in
the case of a triangle, the state with total vorticity two is
realized with three vortices and a single antivortex in
the center of the triangle and preserves $C_3$ symmetry.
Our system with infinite superconducting film
and with magnetic dot on top of it has rotational
$C_{\infty}$ symmetry around the dot center. This
symmetry is reduced in the presence of
vortices and antivortices contrary to the abovementioned example. For the most possible symmetric states with total vorticity zero the
rotational symmetry is either reduced to discrete
$C_{N}$ symmetry (for the case of $N\ge 2$ single
vortices and antivortices) or does not exist at
all (for a vortex--antivortex pair). When two and more (we have checked up to four) vortices and antivortices are present in the zero
vorticity state, vortices are distributed in a
symmetric fashion around the dot center while
antivortices are outside the dot in the same
manner, like homothetically transformed vortices.
In the case of a single vortex--antivortex pair, the dot's center, vortex and antivortex are collinear. Other states with nonzero vorticity
are less symmetric. When the number
of antivortices is five and more, they may form shells around a central
vortex in zero vorticity states.\cite{Mil+03}

The rest of the paper is organized as follows: in
section II we introduce a theoretical model we use
to describe the system. We calculate analytically
the interaction energy between a cylindrical
magnetic dot and vortices in the film. In section III we
determine analytically conditions for the ground state configurations that consist of a single vortex and
a vortex--antivortex pair, as well as separatrices between
those ground states. Positions of vortices are
also found. Magnetic field in whole space and supercurrents in the film are calculated in section IV. Numerical results for the phase
diagram with diversity of vortex--antivortex
configurations as well as their positions are
presented in Section V. Section VI contains numerical estimates of the parameters and conclusions. Some technical details are relegated for the appendix.

\section{Model}

\begin{figure}
\includegraphics[width=0.7\linewidth]{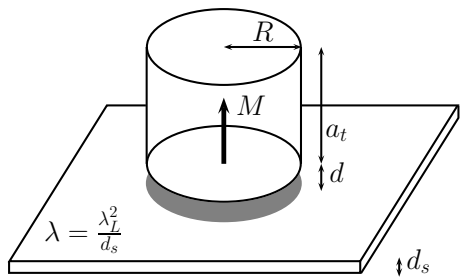}
\caption{Magnetic dot with perpendicular
permanent magnetization $M$ upon infinite
superconducting film.} \label{fig:dot}
\end{figure}

We consider a circular magnetic dot of radius $R$ and
thickness $a_t$ placed above the infinite superconducting
film at distance $d$ with its basis parallel to the film
surface. The dot magnetization $M$ is assumed to be
constant and normal to the film surface, see Fig.~\ref{fig:dot}.
Appart from the supercurrents, the dot may also
induce vortices and antivortices in the film. We will
study these vortex configurations. Since our problem has
many parameters there are many regions in the parameter
space which may have different ground state
configurations(GSC). The GSC here denotes a configuration
of vortices in the film with lowest energy. Among all
possible GSCs a trivial one has no vortices. This is
expected for sufficiently small magnetization. With
increasing $M$ the appearance of vortices is energetically
favorable.

We assume quite generally that our system consists of $N$
vortices with vorticities $n_i$ at positions
$\boldsymbol{\rho}_i$ for a given magnetization $M$. The energy of the
system is
\begin{align}
\label{energy} E_{n_1,\ldots n_N}=\sum_{i=1}^{N}n_i^2
U_v+\sum_{i<j}^{N}n_i n_j
U_{vv}(\rho_{ij})+\sum_{i=1}^{N}n_i U_{mv}(\rho_i),
\end{align}
where
$\rho_{ij}=|\boldsymbol{\rho}_i-\boldsymbol{\rho}_j|$ and
$\rho_i=|\boldsymbol{\rho}_i|$. $U_v$ is the single vortex
energy, $U_{vv}$ is the vortex--vortex interaction, and
$U_{mv}$ is the vortex--magnet interaction. For a given
magnetization of the dot, the system will be in the state
where $E_{n_1,\ldots n_N}$ is minimal.
Expression (\ref{energy}) assumes that the system in the
trivial ground state has zero energy $E_0=0$.

$U_{mv}$ and $U_{vv}$ can be calculated by using the
approach developed in\cite{Erd+02,Erd05a} based on
London--Maxwell equations. Here we only quote final
results. For the interaction energy between the dot and a
vortex with vorticity $n$ placed at distance $\rho$ from
the dot's center we get
\begin{widetext}
\begin{align}\label{Umv}
U_{mv}(\rho,n)=-nMR\phi_0\int_0^{\infty}\dif x
J_0\left(\frac{\rho}{R}x\right)J_1(x)\frac{1}{1+2\frac{\lambda}{R}x}
\frac{\exp\left(-\frac{d}{R}x
\right)-\exp\left(-\frac{d+a_t}{R}x \right)}{x},
\end{align}
where $J_0$ and $J_1$ are the Bessel functions of the first kind. $\lambda$ is the effective penetration depth and is equal $\lambda_L^2/d_s$, where $\lambda_L$ is the London penetration depth and $d_s$ the film thickness. The film thickness is assumed to satisfy $d_s\ll\lambda_L$ and accordingly the magnetic field and currents are uniform through the thickness of the film. Expression (\ref{Umv}) has the following behavior in some regions (we take $d=0$ for simplicity):

\begin{align}\label{Umv_0_approx}
U_{mv}(\rho,n)=-nMa_t\phi_0\frac{R}{4\lambda}\left\{
\begin{array}{l l}
2-\frac{\rho^2}{2R^2},&\hspace{1.5cm}
\rho\ll a_t\ll R\ll\lambda\\
\frac{R}{2a_t}+\frac{R}{a_t}\ln\frac{2a_t}{R}
-\frac{\rho^2}{2Ra_t},
&\hspace{1.5cm} \rho\ll R\ll a_t\ll\lambda\\
\frac{R}{2\lambda}\left(\frac{2\lambda}{\rho}+\ln\frac{\rho}{4\lambda}
+\gamma\right), &\hspace{1.5cm}
a_t,R\ll\rho\ll\lambda\\
\frac{4\lambda^2 R}{\rho^3},&\hspace{1.5cm}
a_t,R\ll\lambda\ll\rho
\end{array} \right.
\end{align}


\end{widetext}
Some technical details for obtaining (\ref{Umv_0_approx}) are presented in the appendix.

The interaction between vortices of vorticities
$n_1$ and $n_2$ separated by a distance $\rho$ is
given by
\begin{align}\label{Uvv}
U_{vv}(\rho)=\frac{n_1
n_2\phi_0^2}{16\pi\lambda}\left[H_0\left(\frac{\rho}{2\lambda}\right)
-Y_0\left(\frac{\rho}{2\lambda}\right)\right],
\end{align}
which has asymptotic forms
\begin{align}
U_{vv}(\rho)=n_1n_2\left\{
\begin{array}{ll}
2U_v, & \text{ } \rho\ll\xi \ll\lambda\\
\frac{\phi_0^2}{8\pi^2\lambda}\left(\ln\frac{4\lambda}{\rho}
-\gamma\right), & \text{ } \xi\ll\rho\ll\lambda\\
\frac{\phi_0^2}{4\pi^2}\frac{1}{\rho}, & \text{ }
\lambda\ll\rho
\end{array} \right.
\end{align}

 In previous expressions $\phi_0$ is the magnetic
flux quantum. $H_0$ is the Struve function of order zero and
$Y_0$ is the Bessel function of the second kind
of order zero.\cite{Abramowitz} $U_v$ is the
single vortex energy\cite{Pea65}
\begin{align}\label{Uv}
U_v=\frac{\phi_0^2}{16\pi^2\lambda}\left(
\ln\frac{4\lambda}{\xi}-\gamma\right),
\end{align}
where $\gamma\approx0.577$ is the Euler constant.
Two vortices which centers are at distances
smaller than $\xi$ are considered in our model as
a double vortex. Its total energy is $4U_v$: from
two single vortices comes $2U_v$ as well as from
their interaction. In the same way a
vortex--antivortex pair at distances smaller than
$\xi$ does not exist, i.e.~it is annihilated.
Here we mention that Eqs.~(\ref{Uvv}) and
(\ref{Uv}) are valid for films with lateral
dimensions much larger than the effective
penetration depth. In the opposite case the
lateral system size enters expressions
(\ref{Uvv}) and (\ref{Uv}) instead of $\lambda$.
This change does not complicate further
considerations and we do not consider such
situation.

In the following we will use energy expression
(\ref{energy}) which will be minimized in the
parameter space and which determines the
structure of vortex configurations in the GS.

\section{Ground states with vortices}

To get further insight into possible GSCs of the
system with respect to different parameters, we
consider asymptotic forms of Eqs.~(\ref{Umv}) and
(\ref{Uvv}). To simplify expressions we will just consider
the case of thin magnetic dots ($a_t<R$) placed at the film
surface ($d=0$). The other cases may be
straightforwardly done having the asymptotic
forms of $U_{mv}$.

Necessary condition for the appearance of an extra vortex
with respect to the trivial GSC is $E_1\le E_0$.
Using (\ref{Umv_0_approx}) in the lowest order in $\lambda/R$ we get
\begin{align}\label{boundary_vortex_thinmagnet}
\mathcal{M}\ge
\frac{2\lambda}{R}.
\end{align}
Here we have introduced $\mathcal{M}=Ma_t\phi_0/U_v$.
This result agrees with one from
Ref.~\cite{Erd+02}. Here we have taken into
account that the strongest attraction energy
between single vortex and magnet occurs at
$\rho=0$. The phase boundary is a linear function of
$\lambda/R$ for $R<\lambda$. For large dot radii $R$, result (\ref{boundary_vortex_thinmagnet}) shows that any nonzero magnetization induces a vortex in the film. This resembles the result for vanishing of the first critical field for thin films in an external homogeneous magnetic field.\cite{Fet+67,Fis80}

Apart from the trivial GSC and the single vortex state,
there are other states for higher $\mathcal{M}$. We will now determine
a portion of the parameter space where a vortex--antivortex
pair appears in the GS. We found that such possibility is
only briefly mentioned in literature,\cite{Sas+00,Mil+02}
or disproved.\cite{Erd05b} We will try to clarify this
issue giving explicit expressions for the phase boundary
and for the vortex and antivortex positions.
\begin{figure}
\includegraphics[width=0.9\linewidth]{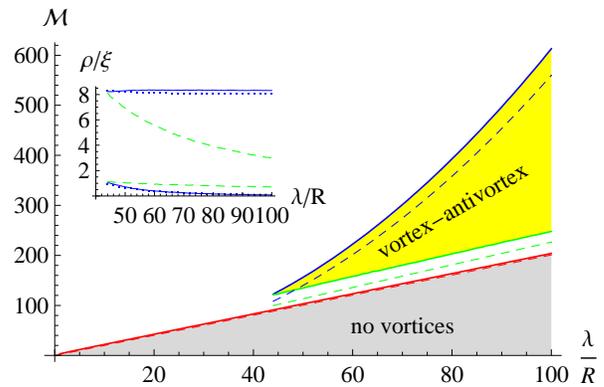}
\caption{(Color online) Phase diagram of a magnetic dot
for $L=6$ and $a_t/R=0.01$: vortex--antivortex state
appears for large enough dot magnetization and large
enough $\lambda/R$. Solid lines are obtained numerically,
while dashed are analytic formulae
(\ref{boundary_vortex_thinmagnet}),
(\ref{boundary_vortex-antivortex_thinmagnet_upper}) and
(\ref{boundary_vortex-antivortex_thinmagnet_lower}) with
$A=1.7$. Inset: upper(lower) curves show
antivortex(vortex) positions. Solid (Dashed) lines
correspond to the upper (lower) phase boundary of the
vortex--antivortex pair. Dotted line are plotted using
analytic formulae (\ref{rho_vortex}) and
(\ref{rho_antivortex}).} \label{1v1av-inset}
\end{figure}

The energy $E_{1,-1}$ of a vortex--antivortex pair
separated by $\rho$ ($\rho\ll2\lambda$) with vortex under
the dot center reads
\begin{align}\label{energy_vortex-antivortex}
\frac{E_{1,-1}(\rho)}{U_v}=&2+\frac{U_{mv}(0,1)}{U_v}+\mathcal{M}\frac{R^2}{4
\lambda
\rho}\\\notag&-\left(\frac{2}{L}+\mathcal{M}\frac{R^2}{8\lambda^2}\right)
\left(\ln\frac{4\lambda}{\rho}-\gamma\right),
\end{align}
where $L$ is the logarithmic factor in the vortex
self energy $L=\ln(4\lambda/\xi)-\gamma$. We may
notice from Eq.~(\ref{energy_vortex-antivortex})
that the main contribution from the magnetic dots
on the antivortex is the energy cost which scales
with the distance from the dot as $1/\rho$, while
the energy gain due to the vortex--antivortex
attraction scales logarithmically which may lead
to a stable potential minimum for some
parameters. For that energy minimum we get that
it occurs at $\rho^*=
2\lambda/(1+\frac{2}{L}\frac{8\lambda^2}{\mathcal{M}R^2})
$ which should be supplemented with the self
consistency condition $\rho^*\ll \lambda$ where
Eq.~(\ref{energy_vortex-antivortex}) is valid and
also $\rho^*>a_t,R$. Implicit equation
$E_{1,-1}(\rho^*)=0$ defines the phase boundary
for the creation of a vortex--antivortex pair. In
addition if $E_{1,-1}(\rho^*)<E_{1}$ and
$E_{1,-1}<0$ are satisfied the vortex--antivortex
pair forms the GS. Condition $E_{1,-1}\le E_{1}$
gives
\begin{align}\label{boundary_vortex-antivortex_thinmagnet_upper}
\mathcal{M}\le\frac{\lambda^2}{R^2}\frac{32}{L\left[
\exp\left(1+\gamma+\frac{L}{2}\right)-2\right]},
\end{align}
which is the upper dashed line for vortex--antivortex
phase boundary in Fig.~\ref{1v1av-inset}. Another phase
boundary we get from the condition that the antivortex is
outside the dot $\rho^*\geq A R$ which gives
\begin{align}\label{boundary_vortex-antivortex_thinmagnet_lower}
\mathcal{M}\geq \frac{8}{L}\frac{\lambda}{R}A
\end{align}
with some number $A$ of order one.

From the last two inequalities we get a condition on
$\lambda/R$ when the vortex--antivortex forms the GS (in
the case $a_t<R, d=0$):
\begin{align}\label{lambdaoverr_thin}
\frac{\lambda}{R}\ge\frac{A}{4}\left[\exp\left(1+\gamma
+\frac{L}{2}\right)-2\right].
\end{align}

In Eq.~(\ref{energy_vortex-antivortex}) we have
assumed that the vortex position is under the dot
$\rho_1=0$ for simplicity. Having in mind
relatively flat magnet--vortex interaction
(\ref{Umv_0_approx}) for $\rho<R$ the system can
gain even more energy allowing  $\rho_1>0$ toward
antivortex, since the energy loss in the
vortex--magnet interaction may be overcompensated
by the vortex--antivortex interaction . A similar
calculation to the previous one gives for the vortex displacement
\begin{align}\label{rho_vortex}
\frac{\rho_1}{R}=\frac{64\lambda^2}{\mathcal{M}^2
R^2 L^2}
\end{align}
while for the antivortex
\begin{align}\label{rho_antivortex}
\frac{\rho}{2\lambda}=\frac{1}{1+\frac{2}{L}\frac{8\lambda^2}
{\mathcal{M}R^2}}-\frac{2R}{\mathcal{M}L\lambda},
\end{align}
and the center of the dot, vortex and antivortex are collinear.
Physically we see that stronger the magnetization of the
dot, the vortex is closer to dot's center, which is expected. What we also see is that the symmetry of a
single vortex GSC is violated for the range of parameters
where the antivortex appears.

\section{Magnetic field}

Magnetic dot on top of a superconducting film induces circular supercurrents in the film. These currents generate magnetic field in and outside the film. Total magnetic field in space is a sum of three terms: due to supercurrents, due to the dot and due to vortices. In this section we calculate the magnetic field using the approach developed in\cite{Erd+02,Erd05a}. The presence of vortices in the film may be observed by measuring the magnetic field and its behavior near the film surface, since vortices change the magnetic field dependence of radial separation from the dot's center.

A vortex of vorticity $n$ produces normal to the film surface (axial) and parallel the film surface (radial) magnetic field which respectively read\cite{Abrikosov}
\begin{align}
&B^{v}_z(\rho,z)=\frac{n\phi_0}{2\pi}\int_0^\infty\dif k\frac{k\exp(-k|z|)}{1+2\lambda k}J_0(k\rho),\\
&B^{v}_{\parallel}(\rho,z)=\frac{n\phi_0}{2\pi}\frac{|z|}{z}\int_0^\infty\dif k\frac{k\exp(-k|z|)}{1+2\lambda k}J_1(k\rho).
\end{align}
The previous expressions at the film surface $z=0$ have the following asymptotic forms:
\begin{align}
\label{Bvz}
& B^v_z(\rho,0)=\frac{n\phi_0}{8\pi\lambda^2}\left\{
\begin{array}{ll}
\frac{2\lambda}{\rho}, & \text{ } \rho\ll\lambda\\
\left(\frac{2\lambda}{\rho}\right)^3, & \text{ } \lambda\ll\rho\\
\end{array} \right.\\
\label{Bvp}
& B^v_{\parallel}(\rho,z\to0)=\frac{n\phi_0}{8\pi\lambda^2}\frac{|z|}{z}\left\{
\begin{array}{ll}
\frac{2\lambda}{\rho}, & \text{ } \rho\ll\lambda\\
\left(\frac{2\lambda}{\rho}\right)^2, & \text{ } \lambda\ll\rho\\
\end{array} \right.
\end{align}
The parallel magnetic field outside the film (remember that in our calculations the film is just in $z=0$ plane) changes the sign going from one side of the film to another. This is due to the fact that the vortex induces currents in the film which circulate around it, and jump in $B^v_{\parallel}$ crossing the film surface is the condition for the jump at the boundaries in electrodynamics due to surface currents.\cite{Jackson} Normal magnetic field is continuous across the film.
The surface current density $K_v$ produced by the vortex is given by
\begin{align}
\mathbf{K}_v=\frac{c}{4\pi}\mathbf{e}_z\times(\mathbf{B}^v_{\parallel}(\rho,0^+)-\mathbf{B}^v_{\parallel}(\rho,0^-)),
\end{align}
where $c$ is the velocity of light.

The magnetic field due to the dot is given by two integrals:
\begin{widetext}
\begin{align}
\label{Bzd}
B^{d}_z(\rho,z)=-2 \pi M R\int_0^\infty\dif kJ_0(k\rho)& J_1(kR)\Big\{\frac{\exp(-k(|z|+d))}{1+2\lambda k}\left[1-\exp(-ka)\right]\\\notag
&+\mathrm{sign}(d-z)\left[1-\exp(-k|d-z|)\right]-\mathrm{sign} (d+a-z)\left[1-\exp(-k|d+a-z|)\right]\Big\},\\
\label{Bdparallel}
B^{d}_{\parallel}(\rho,z)=-2 \pi M R\int_0^\infty\dif kJ_1(k\rho) &J_1(kR)\Big\{\mathrm{sign}(z)\frac{\exp(-k(|z|+d))}{1+2\lambda k}\left[1-\exp(-ka)\right]\\\notag
&+\exp(-k|d-z|)-\exp(-k|d+a-z|\Big\}.
\end{align}
\end{widetext}
Eq.~(\ref{Bzd})(Eq.~(\ref{Bdparallel})) we write as a sum $B^d_{z(\parallel)}=B^{dm}_{z(\parallel)}+B^{df}_{z(\parallel)}$ of fields due to the dot $B^{dm}_{z(\parallel)}$ and due to the supercurrents $B^{df}_{z(\parallel)}$. $B^{dm}_{z(\parallel)}$ is formally defined by setting $\lambda\to\infty$ in Eq.~(\ref{Bzd})(Eq.~(\ref{Bdparallel})) which means the system with the dot and without the film. We evaluate (\ref{Bzd}) and (\ref{Bdparallel}) for $d=0$ and at the film surface. Purely magnetic terms are given by 
\begin{align}
&B^{dm}_z(\rho,0)=-\pi M R^2a_t\frac{1}{\rho^3},\\
&B^{dm}_{\parallel}(\rho,0)=-\pi MR^2 a_t\frac{3a_t}{2\rho^4},
\end{align}
to the leading order for $\rho\gg a_t,R$. The previous result can be understood as the magnetic dipolar field.\cite{Jackson} The part due to supercurrents is
\begin{align}
&B^{df}_{z}(\rho,0)=-\pi MR^2a_t\left\{
\begin{array}{ll}
\frac{1}{4\lambda\rho^2}, & \text{ } a_t,R\ll\rho\ll\lambda\\
\mathcal{O}\left(\frac{1}{\rho^4}\right), & \text{ } a_t,R\ll \lambda\ll\rho\\
\end{array} \right.\\
&B^{df}_{\parallel}(\rho,z\to0)=-\pi MR^2a_t\frac{|z|}{z}\left\{
\begin{array}{ll}
\frac{1}{\lambda\rho^2}, & \text{ } a_t,R\ll\rho\ll\lambda\\
\frac{6\lambda}{\rho^4}, & \text{ } a_t,R\ll \lambda\ll\rho\\
\end{array} \right.
\end{align}
Again parallel field $B^{df}_{\parallel}$ jumps across the film surface due to surface currents. The surface current density $K_m$ in the film due to the presence of the dot is given by
\begin{align}
\mathbf{K}_m=\frac{c}{4\pi}\mathbf{e}_z\times(\mathbf{B}^{df}_{\parallel}(\rho,0^+)-\mathbf{B}^{df}_{\parallel}(\rho,0^-)).
\end{align}

As first proposed in Ref.~\cite{Erd+02} the presence of a single vortex in the film can be proved by observing the change of sign of the total field $B_z$ near the film. We can now calculate that it happens when $B^d_z(\rho_{sz},0)+B^v_z(\rho_{sz},0)=0$ or at $\rho_{sz}=2\pi\sqrt{MR^2a_t\lambda/\phi_0}$ provided a single vortex appears in the GS. Using the condition for the single vortex (\ref{boundary_vortex_thinmagnet}) we get $\rho_{sz}\approx\sqrt{LR\lambda/2}$. We can also formulate a similar condition for the presence of a single vortex but for the parallel field, which also changes sign for at distance $\rho_{s\parallel}$ from the dot defined by $B^d_{\parallel}(\rho_{s\parallel},0)+B^v_{\parallel}(\rho_{s\parallel},0)=0$. The solution of the resulting cubic equation is $\rho_{s\parallel}\approx (6\pi^2M R^2a_t^2\lambda/\phi_0)^{1/3}$ above the film ($z\to 0^{+}$), which using (\ref{boundary_vortex_thinmagnet}) becomes $\rho_{s\parallel}\approx(3LRa_t\lambda/4)^{1/3}$. 

\begin{figure}
\includegraphics[width=0.95\linewidth]{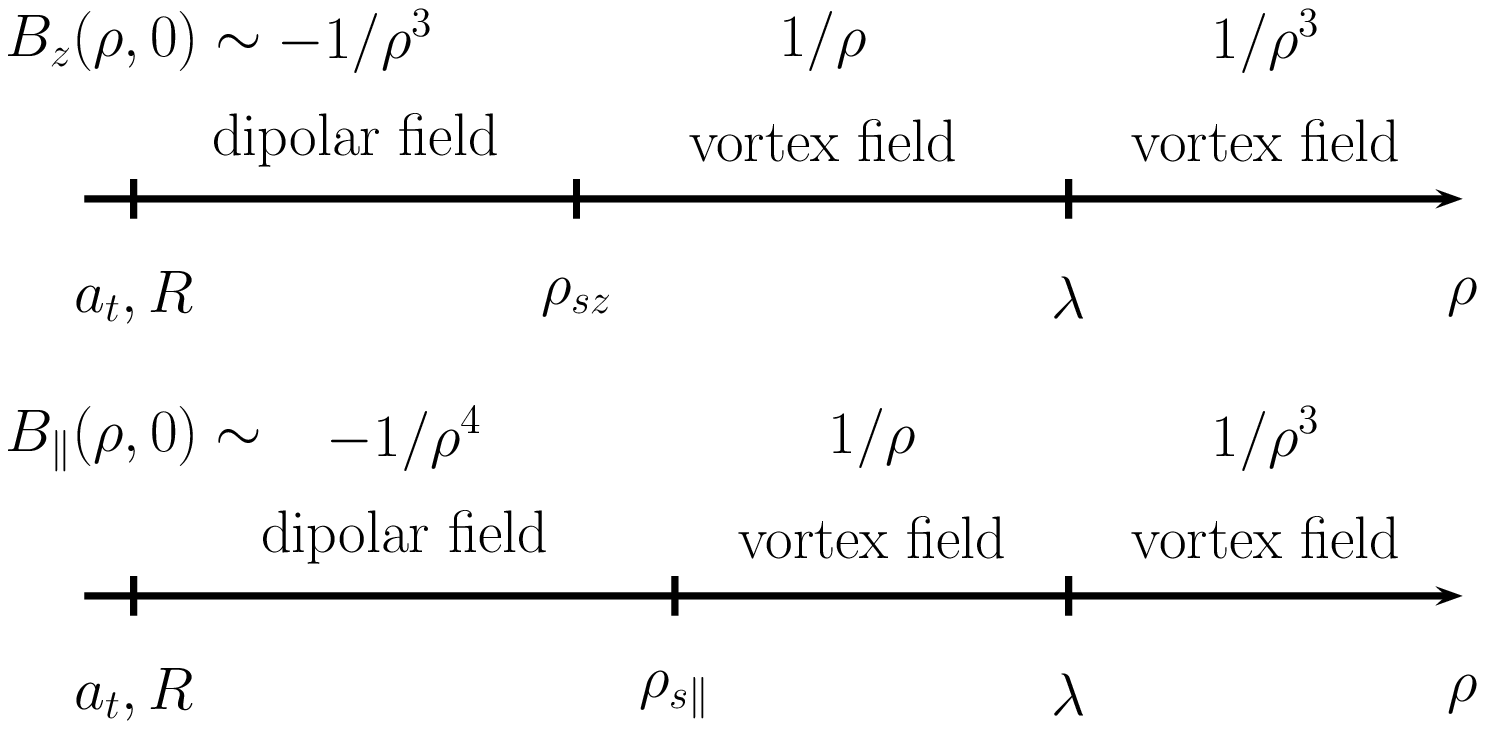}
\caption{Behavior of the magnetic field near the upper film surface. Magnetization of the dot is assumed to be directed as in Fig.~\ref{fig:dot} such that a single vortex appears under the dot. The magnetic field changes the sign at at some distance from the dot due to the presence of the vortex.}
\label{fig:fields}
\end{figure}

Qualitatively different behavior of the normal and parallel magnetic field is summarized in Fig.~\ref{fig:fields} when a single vortex is present in the GS. Sufficiently close to the dot dominates the dipolar field from the dot, while at larger distances the vortex part of the field is a leading term.

So far we have only considered a single vortex under the dot. Since there may be many other vortex--antivortex states in the film, the magnetic field (vortex part only) of such configurations behaves differently than in (\ref{Bvz}) and (\ref{Bvp}) and will get angular dependence. We do not analyze this case. Again, the total magnetic field close enough to the dot will be dominated by the dipolar field from the dot, while at larger distances the anisotropic vortex part of the magnetic field prevails. One should be able to use the measurement of magnetic field near the film for detection of many--vortex states bounded by the dot.

\section{Numerical study of ground states with low number of vortices}

In section III we have shown analytically a
possibility of having a vortex--antivortex pair in the GS.
That was the simplest GSC with antivortex. Certainly there
are other more complicated states for a range of
parameters. Analytic study of these states is in principle
straightforward using already introduced expressions, but
tedious. In this section we study numerically GSCs. For a
given configurations uniquely determined by
$n_1,n_2,\ldots$ we use expression (\ref{energy}) to
obtain the energy minimum and positions of vortices and
antivortices. The GSC for given $\lambda/R$, $a_t/R$, $L$
and $\mathcal{M}$ has the energy minimum over of all
possible vortex--antivortex configurations. In our
calculations we took into account just states with low
numbers of vortices, since they cover significant part of
the parameter space as well as the other states are
computationally demanding. Dot--magnet distance is set to
$d=0$. We have examined all states which
have up to four vortices with antivortices, while for the
states with five vortices we took into account vortices
without antivortices. This is certainly not correct since
antivortices appears in states with five vortices as well,
but such states are located in a particular region of the
parameter space and do not affect states with up to three
vortices (see further in the text).

States with six and more vortices are not taken
into account. This is not crucial for our study
since the vortex state with vorticity $n$ will
appear in the phase diagram for
$\mathcal{M}\ge2n\lambda/R$ which is
either for large magnetization or large
magnetic dots (with respect to $\lambda$). On the
other hand line with $n$ single vortices (and
antivortices) is moved a little bit toward
smaller $\mathcal{M}$ for a given $\lambda/R$.

\begin{figure}
\includegraphics[width=1\linewidth]{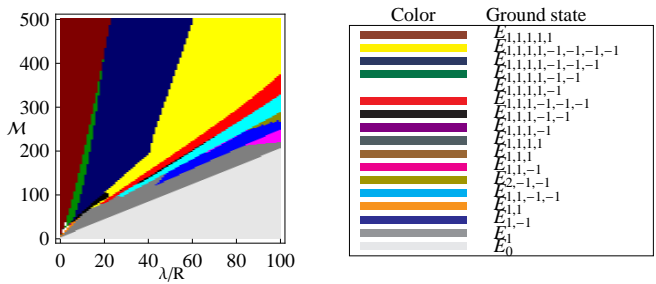}
\caption{(Color online) Phase diagram of a magnetic dot
for $L=6$ and $a_t/R=0.01$ with different
vortex--antivortex configurations.}
\label{fig-gs-full-001}
\end{figure}

\begin{figure}
\includegraphics[width=1\linewidth]{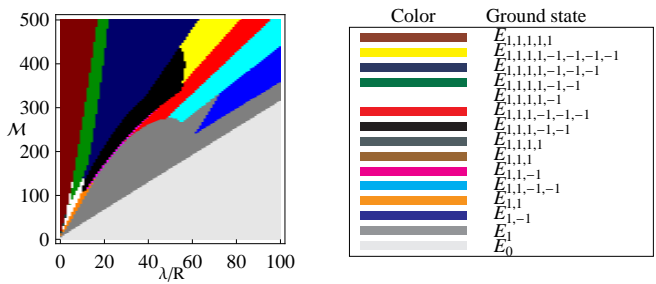}
\caption{(Color online) Phase diagram of a magnetic dot
for $L=6$ and $a_t/R=1$ with different vortex--antivortex
configurations.} \label{fig-gs-full-1}
\end{figure}
In Fig.~\ref{fig-gs-full-001} we show the GSC of the
magnetic dot with $a_t/R=0.01$ and $L=6$. For low
$\mathcal{M}$ there are no vortices in the GS for any
$\lambda/R$. With increasing $\mathcal{M}$ state
$E_1$ appears. Further increase of $\mathcal{M}$ is
ultimately followed with the antivortex appearance in the
GS. To be concrete let us consider $\lambda/R=50$. After
$E_1$, state with antivortex $E_{1,-1}$ appears which is
the GSC for some region of $\mathcal{M}$. Then states
$E_{1,1,-1,-1}$, $E_{1,1,1,-1,-1,-1}$, $E_{1,1,1,-1,-1}$,
$E_{1,1,1,1,-1,-1,-1,-1}$ appear for larger $\mathcal{M}$.
Further states might have included states with five
vortices with antivortices if we had taken them into
account. We see that states that evolve from each other
have vortex and/or antivortex more/less with respect to
its neighbors. We believe that this should mean that
states with three vortices will not be affected by the
states with five vortices.

Also the net vorticity of the GSCs is in general
small ($0,\pm 1$), just near the origin it is
higher. However for small $\lambda/R$ the dot
radius is pretty large for realistic films and
further analysis with finer resolution of
$\lambda/R$ is necessary. We are here mainly
concentrated in the region where $R<\lambda$ since it is the most interesting experimentally, see the last section for some numerical values of parameters.

\begin{figure}
\includegraphics[width=0.8\linewidth]{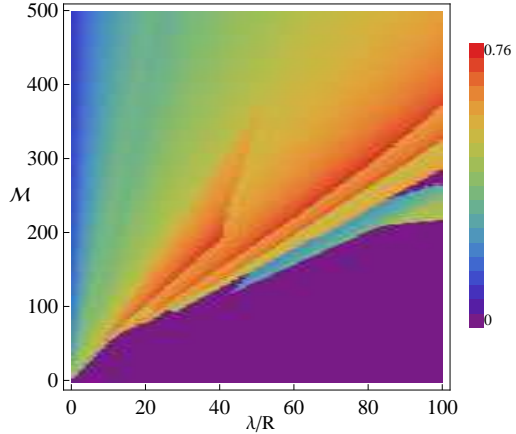}
\caption{(Color online) Vortex distances (in units of dot
radius $R$) from the dot center for $L=6$ and $a_t/R=0.01$
which correspond to a particular ground state
configuration given in FIG.~\ref{fig-gs-full-001}.}
\label{fig-vdistances}
\end{figure}
\begin{figure}
\includegraphics[width=0.8\linewidth]{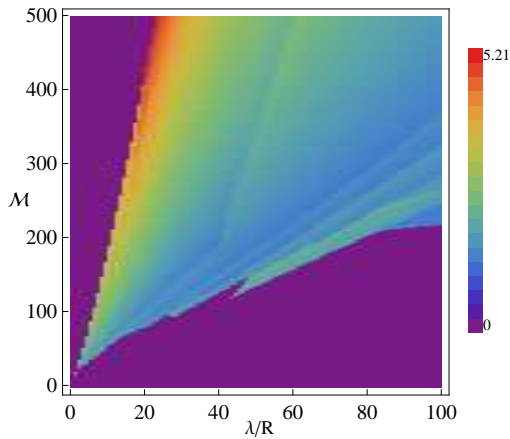}
\caption{(Color online) Antivortex distances (in units of
dot radius $R$) from the dot center for $L=6$ and
$a_t/R=0.01$ which correspond to a particular ground state
configuration given in FIG.~\ref{fig-gs-full-001}.}
\label{fig-avdistances}
\end{figure}

The distance between the center of the dot and
vortices and antivortices are shown in
Fig.~\ref{fig-vdistances} and
Fig.~\ref{fig-avdistances} respectively. We see
that they abruptly change at boundaries between
different GSCs. We also see that antivortices
when are present are at distances of a few $R$.
Inside the same GSC for fixed $\lambda/R$ with
increasing $\mathcal{M}$ the antivortices spread,
while vortices shrink. This is plausible: very
large magnetization of the dot would expel
antivortices to very large distances since the
interaction energy cannot be then compensated by
the attractive vortex--antivortex attraction. On
the other hand vortices are attracted at smaller
distances toward the potential minimum of the
vortex--magnet interaction. These results are
also confirmed by analytic formulae
(\ref{rho_vortex}) and (\ref{rho_antivortex})
from the study of a single pair.

We also find that in states with zero total
vorticity and two,three and four vortices,
positions of antivortices are obtained as
homothetically transformed positions of vortices,
and they form line, equilateral triangle and
square, respectively.

\begin{figure}
\includegraphics[width=0.8\linewidth]{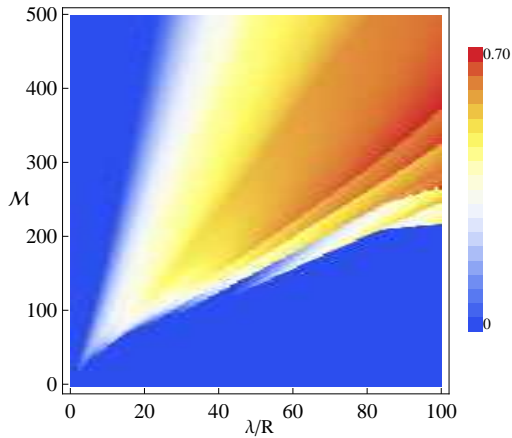}
\caption{(Color online) Relative energy change in the
ground state when antivortices are included for $L=6$ and
$a_t/R=0.01$.} \label{fig-energychange}
\end{figure}

\begin{figure}
\includegraphics[width=1\linewidth]{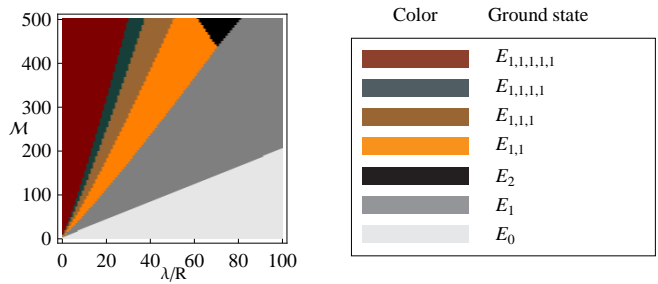}
\caption{(Color online) Phase diagram of a magnetic dot
for $L=6$ and $a_t/R=0.01$ without antivortices.}
\label{fig-gs-justv-001}
\end{figure}

The GSCs without antivortices
are shown in Fig.~\ref{fig-gs-justv-001}. Relative energy difference
due to the presence of antivortices is shown in
Fig.~\ref{fig-energychange}. The relative
energy gain is quite significant which means that states
without and with antivortices have quite different
energies.


We see from Fig.~\ref{fig-energychange} that for smaller
magnetization the vortex states with single vortices appear,
while state with double vortex $E_2$ is present for higher
$\mathcal{M}$. Simple analytical check gives that
condition $E_{1,1}\le E_{2}$ translates into
\begin{align}\label{giant-single}
\mathcal{M}\le\frac{1}{L}\exp\left(2L+1-2\gamma\right)\frac{R}{\lambda},
\end{align}
which for $L=6$ becomes $\mathcal{M}\le 23200R/\lambda$.
That also gives one explanation why the only ground state
with giant vortices in Fig.~\ref{fig-gs-full-001} is
$E_{2,-1,-1}$: it appears around $\lambda/R=100,
\mathcal{M}=250$ in agreement with rough estimate
(\ref{giant-single}). The other states with giant vortices
may appear only for higher $\mathcal{M}\lambda/R$ ratios.

For comparison with thicker magnetic dots we have
calculated the GS for $a_t/R=1$ and they are shown in
Fig.~\ref{fig-gs-full-1}. The single vortex appears now
for larger $\mathcal{M}$, which is obvious from
(\ref{boundary_vortex_thinmagnet}). State $E_{1,-1}$ now
occurs for larger $\lambda/R$. However a global picture
with diversity of vortex--antivortex states again holds.
However GSC here are shifted toward higher $\lambda/R$ and
$\mathcal{M}$ values with respect to the corresponding
states for thinner dots. We may interpret this as a fact
that thicker magnets are less efficient in exciting
vortices which is due to the larger extent of the dot with
respect to the film surface. Very thick dots ($a_t\gg R$)
interact with vortices weakly, see two cases in formula
(\ref{Umv_0_approx}).

Now we comment about the applicability of the
London approximation for our case.
The dot radius over the coherence length $\xi$ as
a function $\lambda/R$ is given as
$R/\xi=R\exp(L+\gamma)/(4\lambda)$. We may
conclude that for $\lambda/R<100$ and $L=6$ the
dot radius is always larger than the coherence
length. On the other hand a necessary condition
for the London approach to be valid is that all
lengths are (much) larger than the coherence
length. We see that this is satisfied better as
$\lambda/R$ approaches smaller values. We expect
that our results with different GSCs  are valid
even qualitatively for $\lambda/R\lesssim 50$,
while the general picture with antivortices and
low vorticity GSC holds even for smaller dot
radii. However it is expected that for smaller
$R/\xi$ ratios, giant vortices are favorable. We
have already mentioned such tendency.

Let us mention that a certainly better numerical
way would be the nonlinear Ginzburg--Landau
theory. For the
case of mesoscopic superconducting discs comparison  of these two methods  shows that both methods give similar results for
$R=6\xi$ and up to five vortices.\cite{Baelus+04}

\section{Discussions and conclusions}

Let us consider numerical values of parameters
for realistic systems. For thin Nb films close to
the critical temperature $T/T_c=0.98$ values for
the London penetration depth and the coherence
length are $\lambda_L\cong560\ \mathrm{nm}$ and
$\xi\cong58\ \mathrm{nm}$.\cite{Hoffmann+00} Then
$L=6$ corresponds to a film of thickness
$d_s\cong30\ \mathrm{nm}$ with the effective penetration depth $\lambda\cong 10.5\
\mathrm{\mu m}$. Condition for $E_{1,-1}$ state
(\ref{lambdaoverr_thin}) can be rewritten as
$R<\lambda_L\sqrt{\xi/(9d_s)}$ or in our case
$R<260\ \mathrm{nm}$. The condition for the
applicability of the London theory $\xi<R$ is
still satisfied, so the results should be valid.
Larger magnetic dots may have other
vortex--antivortex configurations, and the London
theory is expected to be applicable.

In this paper we have considered infinite films.
We expect this not to be a severe limitation as
soon as the dot radius is much smaller than the
system size since the antivortices when appear
are at distances of the order of $R$ and the
boundaries of the film should not affect it.
For films with lateral dimensions smaller than $\lambda$ the single vortex energy, Eq.~(\ref{Uv}) will have under the logarithm the lateral dimension instead of $4\lambda$, and for small dot radii essentially the same story should be repeated, just with new $U_v$.

To conclude, in this paper we have considered a thin superconducting film
with a cylindrical magnetic dot with permanent magnetization
upon it. Inside the Maxwell--London approach we have
calculated the vortex--magnet interaction as well its
asymptotic limits. Using these results we have shown
analytically that a vortex--antivortex pair appears in the
ground state of the system for some range of dot's radii
and its magnetization. Necessary magnetization for that is
comparable to the magnetization for the appearance of a single vortex. Magnetic field everywhere in space is also calculated. Near the film surface, the magnetic field has different scaling forms with the distance from the dot center. This fact may be used for the experimental detection of vortices. In addition to that we have calculated numerically the phase diagram with up to four vortices with antivortices.

\subsection*{Acknowledgments}
This work is financially supported by the DFG under the grant
NA222/5-2 and partly by the DOE under the grant DE-FG02-06ER46278.
The author wishes to thank Prof. V. Pokrovsky for many fruitful discussions and Prof. T. Nattermann for his support.

\section{Appendix}

In this appendix we calculate asymptotically the integrals that appears in the interaction energy (\ref{Umv}). We consider an integral of the form
\begin{align}
\label{Iabc}
I(a,b,c)=\int_0^\infty\dif x J_0(a x)J_1(x)\frac{\exp(-c x)}{x(1+2b x)}.
\end{align}

In the limit $a\gg c$ and $a\gg 1$ this integral will be cut by the oscillations of the Bessel functions and the main contribution comes from the region $x<1/a\ll 1$. Then we can expand $J_1(x)\exp(-c x)\approx x(1-c x)/2$. Using the tabulated integrals\cite{Prudnikov}
\begin{align}
&\int_0^\infty\dif x\frac{J_0(\alpha x)}{1+x}=\frac{\pi}{2}\left[H_0(\alpha)-Y_0(\alpha)\right],\\
&\int_0^\infty\dif xJ_0(\alpha x)=\frac{1}{\alpha},
\end{align}
we easily get
\begin{align}
I(a,b,c)=\frac{2b+c}{8b^2}\frac{\pi}{2}\left[ H_0\left(\frac{a}{2b}\right)-Y_0\left(\frac{a}{2b}\right)\right] -\frac{c}{4ab}.
\end{align}
Using the expansion\cite{Abramowitz}
\begin{align}
\frac{\pi}{2}\left[Y_0(x)-H_0(x)\right]=\left\{
\begin{array}{l l}
{\gamma}+\log \frac{x}{2}-x,& \text{ }x\ll 1\\
-x^{-1}+x^{-3},& \text{ } x\gg 1
\end{array}\right.
\end{align}
where $\gamma\approx0.577$ is the Euler constant, one can further simplify the asymptotic expressions for $I(a,b,c)$.

In the limit $a \ll c$ and $a\ll 1$ integral (\ref{Iabc}) is cut by the exponential function at $x\approx 1/c$, which means the argument of $J_0$ function $ax\approx a/c\ll 1$ and we expand $J(0,ax)\approx 1-a^2 x^2/4$. We get
\begin{align}
I(a,b,c)=\int_0^\infty\dif x J_1(x)\exp(-c x)\\\notag\times\left(-\frac{a^2}{8b}+\frac{1}{x}
+\frac{\frac{a^2}{8b}-2b}{1+2b x}\right).
\end{align}
Then using the tabulated integrals\cite{Prudnikov}
\begin{align}
&\int_0^\infty\dif x\frac{J_1(\alpha x)}{1+x}=1+\frac{1}{\alpha}+\frac{\pi}{2} \left[Y_1(\alpha)-H_1(\alpha)\right],\\
&\int_0^\infty\dif xJ_1(\alpha x)\frac{\exp(-\gamma x)}{x}=-\gamma+\sqrt{1+\gamma^2},\\
&\int_0^\infty\dif xJ_1(\alpha x)\exp(-\gamma x)=1-\frac{\gamma}{\sqrt{1+\gamma^2}},
\end{align}
and the expansion for $x\ll 1$\cite{Abramowitz}
\begin{align}
\frac{\pi}{2}\left[H_1(x)-Y_1(x)\right]=\frac{1}{x}+\frac{x}{4}\left( 1-2\gamma+\log\frac{4}{x^2}\right)
\end{align}
we get for $a\ll c\ll 1\ll b$
\begin{align}
I(a,b,c)&=\frac{1}{8b}\bigg(1-2\gamma+c(a^2-4)\\\notag &
-a^2\sqrt{1+c^2}+\log\frac{16}{b^2}\bigg)+c-\frac{c}{\sqrt{1+c^2}}. 
\end{align}
I the above expressions $J_0$ and $J_1$ are the Bessel functions of the first kind, $Y_0$ and $Y_1$ are the Bessel functions of the second kind, while $H_0$ and $H_1$ are the Struve functions of order zero and one, respectively.\cite{Abramowitz}


\end{document}